\documentclass[aps,preprint,12pt,floatfix,nofootinbib]{revtex4-2}

\usepackage{amsmath}
\usepackage{amssymb}
\usepackage{xcolor}
\usepackage{graphicx}
\usepackage{dcolumn}
\usepackage{float}
\usepackage{bm}
\usepackage{caption}
\usepackage{subfig}
\usepackage{epstopdf}
\usepackage{epsfig}

\usepackage[italicdiff]{physics}
\usepackage{hyperref}

\hypersetup{colorlinks=true,
            allcolors=blue}

\newcommand{\orcidicon}[1]{\href{https://orcid.org/#1}{\includegraphics[height=\fontcharht\font`\B]{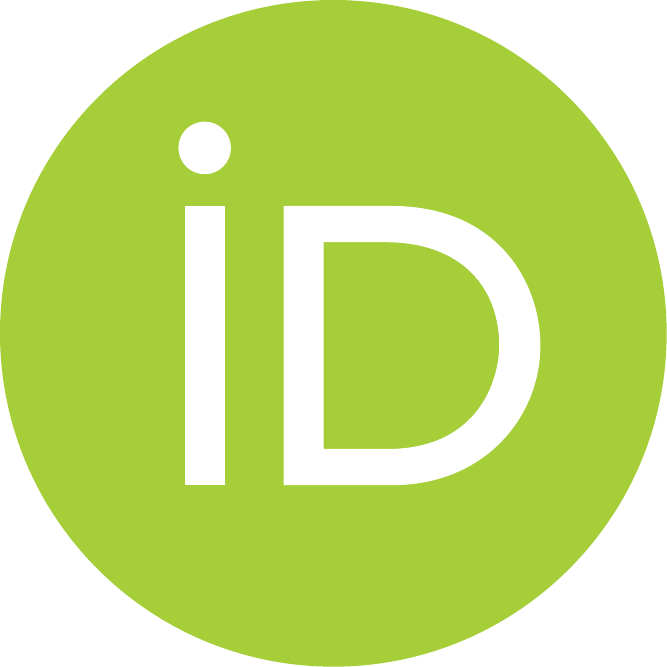}}}

\newcommand\blfootnote[1]{%
  \begingroup
  \renewcommand\thefootnote{}\footnote{#1}%
  \addtocounter{footnote}{-1}%
  \endgroup
}

\begin{document}

\title{Closest approach of a quantum projectile}

\author{A Kumar$^{1,2}$\,\orcidicon{0000-0003-3639-6468}}
\author{T Krisnanda$^2$\,\orcidicon{0000-0002-6360-5627}}
\author{P Arumugam$^1$\,\orcidicon{0000-0001-9624-8024}}
\author{T Paterek$^{2,3,4}$\,\orcidicon{0000-0002-8490-3156}}

\affiliation{$^\mathit{1}$Department of Physics, Indian Institute of Technology Roorkee, Roorkee 247667, India}
\affiliation{$^\mathit{2}$School of Physical and Mathematical Sciences \\ Nanyang Technological University, Singapore 637371, Singapore}
\affiliation{$^\mathit{3}$Institute of Theoretical Physics and Astrophysics, Faculty of Mathematics, Physics and Informatics, University of Gda\'{n}sk, 80-308 Gda\'{n}sk, Poland}
\affiliation{$^\mathit{4}$MajuLab, International Joint Research Unit UMI 3654 \\ CNRS, Universit\'{e} C\^{o}te d’Azur, Sorbonne Universit\'{e}, National University of Singapore, Nanyang Technological University}

\blfootnote{A Kumar: \textcolor{blue}{ akumar18@ph.iitr.ac.in}}

\begin{abstract}
We consider the simplest case of Rutherford scattering, i.e. the head-on collision, where the projectile is treated quantum mechanically. The convexity of repulsive Coulomb force invokes a disagreement between the Ehrenfest's and Hamilton's dynamics. We show that the quantum projectile cannot approach as close as the corresponding classical one, and that the average distance of closest approach depends on the position spread of the wave function describing the projectile.
\end{abstract}

\maketitle

\section{Introduction}

The classical dynamics of a particle of mass $m$, in a potential $V(x)$, is governed by the Hamilton's equations:
\begin{equation}
\dv{t}x = \frac{p}{m}, \hspace{10mm} \dv{t}p = - V'(x),
\end{equation}
where $V' = dV/dx$, and $p$ is the momentum. On the other hand, the average quantum dynamics is governed by Ehrenfest's equations:
\begin{equation}
\dv{t}\ev{x} = \frac{\ev{p}}{m}, \hspace{5mm} \dv{t}\ev{p} = -\ev{V'(x)}.
\end{equation}
The two sets of equations are very similar, but not the same. They are exactly the same under the condition
\begin{equation}
\ev{V'(x)} = V'(\ev{x}),
\end{equation}
which is true only when the potential is at most quadratic in $x$. None of the fundamental interactions satisfy this criteria and accordingly even the average quantum dynamics they generate may differ from the classical trajectories.

Here we vividly illustrate this with a simple case of Rutherford scattering, i.e., the head-on collision~\cite{vyas-arXiv:2011.06470}. Following the seminal approach we model projectile-target interactions as due to Coulomb force only. Differently from the original approach we assign a wave function to the projectile, i.e. the quantum projectile (alpha particle) with wave function $\psi(x)$ moves in the Coulomb potential generated by gold nucleus. The Coulombic potential varies with distance like $1/x$, which implies
\begin{equation}
\ev{V'(x)} \sim\ev{\frac{1}{x^2}}, \hspace{10mm} \text{and} \hspace{10mm} V'(\ev{x}) \sim \frac{1}{\ev{x}^2}.
\end{equation}
Clearly, $\ev{V'(x)} \neq V'(\ev{x})$ unless $\psi(x) \equiv \delta(x-x')$. This leads to Hamilton's solutions for $x(t)$ and $p(t)$ which do not match with the Ehrenfest's solutions for $\ev{x(t)}$ and $\ev{p(t)}$~\cite{book_QM_MaxJammer,book_QM_BCHall,vyas-arXiv:2011.06470}. We study the dynamics of the projectile and investigate the quantum mechanical features of the event. In the following section, we outline the way of calculating the time evolution of the wave function.

\section{Cayley's form of time evolution operator} \label{sec:CayleyMethod}

For time-independent Hamiltonians, the time evolution of a quantum state $\psi(x,t)$ is governed by the evolution operator $U(t+\Delta t,t) = \exp{-iH\Delta t/\hbar}$, where $\Delta t$ is the time step and $H=-\frac{\hbar^2}{2m}\pdv[2]{x} + V(x)$ is the Hamiltonian. Any truncation in the binomial expansion of $U$ leads to a loss of unitarity, and consequently a change in the norm of wave function over time. To circumvent this problem we use the Cayley's form~\cite{JCP2794,JCP041101,book_CompPhys_FJVesely}:
\begin{equation}
U(t+\Delta t,t) \approx \left( 1 + i\frac{H\Delta t}{2\hbar} \right)^{-1} \left( 1 - i\frac{H\Delta t}{2\hbar} \right).
\end{equation}
Therefore, $\psi(x,t)$ and $\psi(x,t+\Delta t)$ are related by the Crank-Nicolson relation:
\begin{equation}
\left( 1 + i\frac{ H\Delta t}{2\hbar} \right) \psi(x,t+\Delta t) = \left( 1 - i\frac{ H\Delta t}{2\hbar} \right) \psi(x,t).
\end{equation}
Hence, the idea is to evolve the system half of the time step forward-in-time, and half of the time step by the inverse of backward-in-time evolution. We approximate the second derivative in the Hamiltonian by the three-point central difference formula to write
\begin{equation} \label{TLSE}
\begin{split}
\psi_j^{n+1} + \frac{i \Delta t}{2\hbar} \left[ -\frac{\hbar^2}{2m} \left( \frac{\psi_{j+1}^{n+1} - 2\psi_{j}^{n+1} + \psi_{j-1}^{n+1}}{dx^2} \right) + V_j  \psi_j^{n+1} \right] \\    = \psi_j^{n} - \frac{i \Delta t}{2\hbar} \left[ -\frac{\hbar^2}{2m} \left( \frac{\psi_{j+1}^{n} - 2\psi_{j}^{n} + \psi_{j-1}^{n}}{dx^2} \right) + V_j  \psi_j^{n} \right],
\end{split}
\end{equation}
where $f_j^n \equiv f(x_j,t_n)$, $\Delta x=x_{j+1}-x_j$ is the grid size, and $\Delta t=t_{n+1}-t_n$ is the time step. Eq.~(\ref{TLSE}) represents a tridiagonal system of linear equations for the unknown wave function at the next time step ($\psi^{n+1}$). The solution is calculated by performing an LU-factorisation of the tridiagonal matrix on the left, followed by forward- and backward substitutions of the vector on the right~\cite{CoP_PeterPuschnig}.

\section{Results and discussions}

The initial wave function, describing the alpha particle at $t$ = 0, is given by
\begin{equation}
\psi(x,0) = \sqrt{\frac{1}{\sigma\sqrt{2\pi}}} \exp{-\frac{(x-x_0)^2}{4\sigma^2} + i\frac{p_0}{\hbar}(x-x_0)},
\end{equation}
where $x_0$, $p_0$ and $\sigma$ are the average position, average momentum and the position spread, respectively. The Coulombic potential experienced by the particle is (the nucleus is at $x=0$):
\begin{equation}
V(x) = Z_{He}Z_{Au}\frac{\alpha \hbar c}{|x|},
\end{equation}
where $Z_{He}=2$ and $Z_{Au}=79$ are the atomic numbers of helium and gold, respectively, $\hbar c = 197.3269631$ MeV fm, and $\alpha = 1/137.035999679$ is the fine-structure constant characterizing the electromagnetic interaction. In the classical case, we consider alpha particles are shot from $x = -500$ fm with a momentum of $386.13$ MeV/c (classically equivalent to a kinetic energy of $20$ MeV). The time evolution is calculated by integrating the equation of motion with appropriate boundary conditions. In the quantum case, we consider Gaussian wave packets centered about $x_0 = -500$ fm with a width $\sigma$ and an average momentum of $p_0 = 386.13$ MeV/c. The time evolution is calculated using the Cayley's operator as discussed in Sec.~\ref{sec:CayleyMethod}.

\begin{figure}[h]
\centering
\subfloat[Trajectories.]{\includegraphics[width=0.475\textwidth]{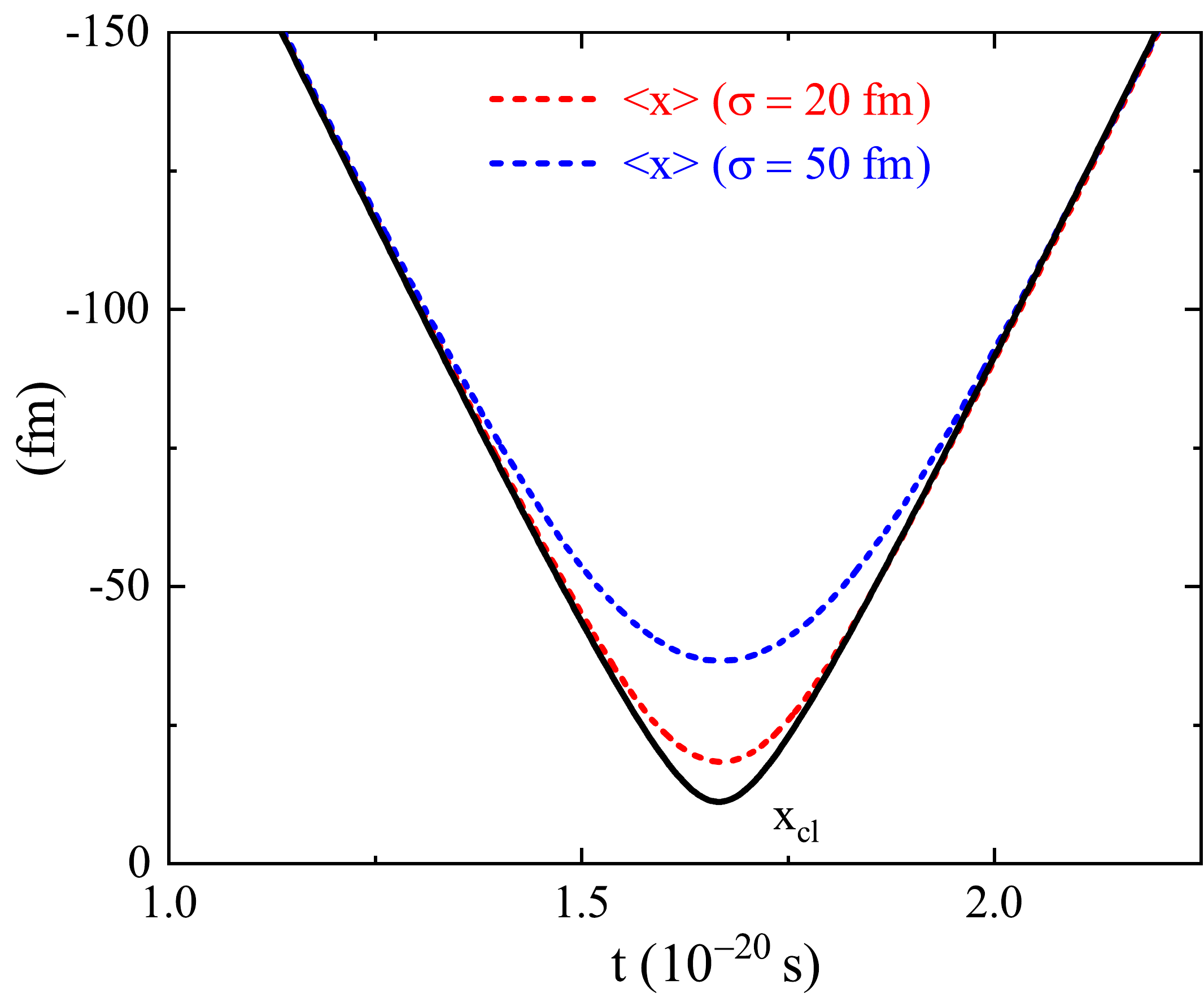} \label{fig:trajectory}}
\hfill
\subfloat[Forces.]{\includegraphics[width=0.475\textwidth]{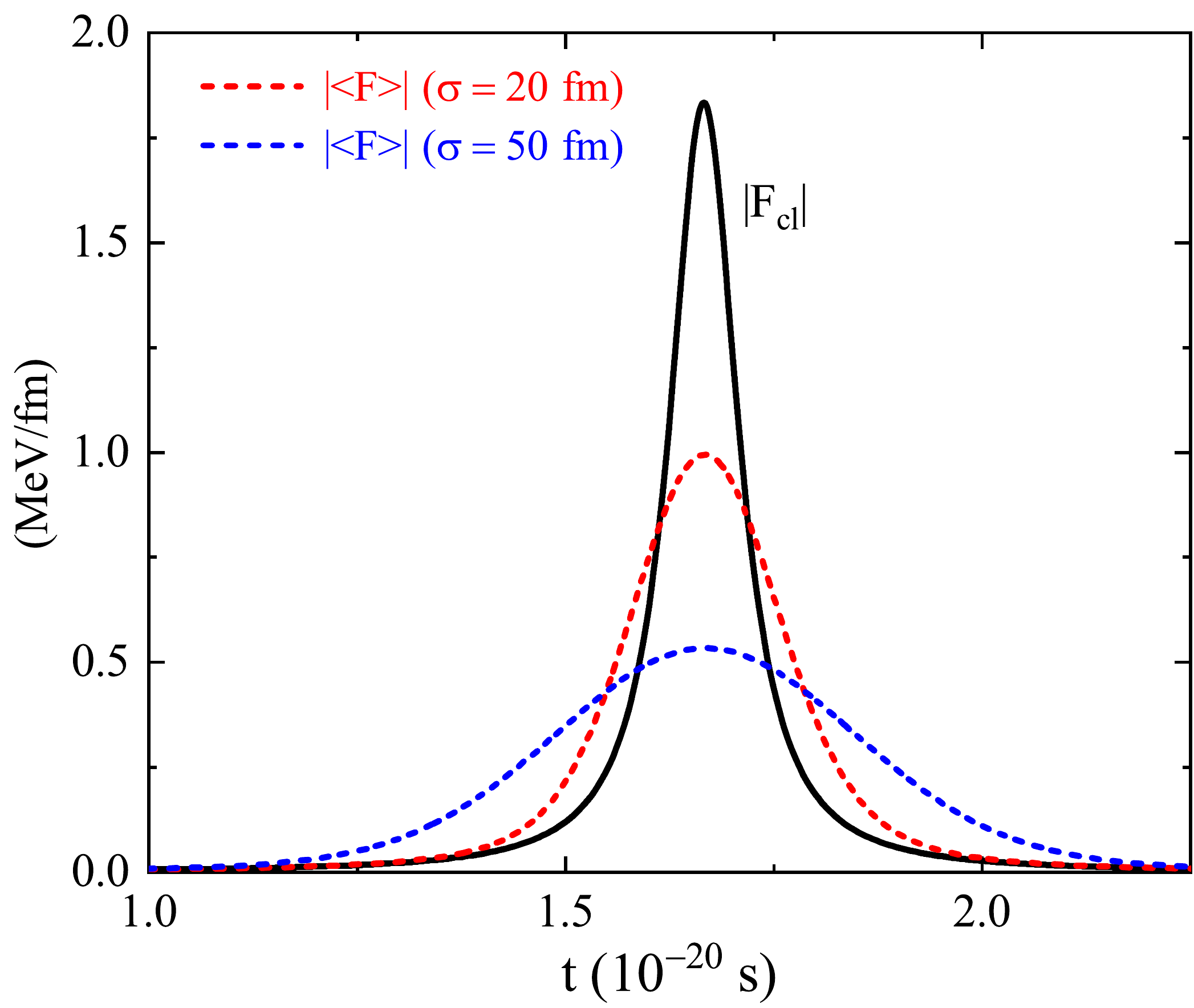} \label{fig:force}}
\caption{Comparison of classical and quantal trajectories of alpha particle shot at gold nucleus fixed at the origin. $x_{cl}$ and $F_{cl}$ denote the position and the force acting on alpha particle, respectively, in the classical case. $\ev{x}$ and $\ev{F}$ are the average position and the average force, respectively, experienced according to the quantum theory. $\sigma$ is the initial width of the Gaussian wave packet. Initial position $= -500$ fm, and initial momentum $= 386.13$ MeV/c.}%
\label{fig:trajecforce}%
\end{figure}
Fig.~\ref{fig:trajectory} compares the classical path with the quantal trajectories for $\sigma = 20$ and $50$ fm. $\ev{x(t)}$ starts deviating from $x_{cl}(t)$ as soon as the projectile advances towards the target, and the degree of deviation increases with the position spread of the wave function. This is a striking implication of the convexity of the repulsive Coulomb force. Since initially $\ev{x(0)} = x_{cl}(0) \equiv x_0$, we can use the Jensen's inequality~\cite{ActaMath175} to write~\cite{vyas-arXiv:2011.06470}:
\begin{equation}
|\ev{F}| \sim \ev{\frac{1}{x^2(0)}} \geq \frac{1}{\ev{x(0)}^2} = \frac{1}{x_0^2} =  \frac{1}{x_{cl}^2(0)} \sim |F_{cl}|.
\label{inequality_teq0}
\end{equation}
The average quantum force is larger than the force experienced in the classical case. This is also clearly visible in Fig.~\ref{fig:force}. As a result, the quantum particle moves slower than its classical counterpart, i.e. at later times $t > 0$, we have
$\ev{x(t)} < x_{cl}(t)$. Recall from Fig.~\ref{fig:trajectory} that both of these values are strictly negative and hence $\ev{x(t)}^2 > x_{cl}^2(t)$. At later times we therefore have~\cite{vyas-arXiv:2011.06470}
\begin{equation}
|\ev{F}| \sim \ev{\frac{1}{x^2(t)}} \geq \frac{1}{\ev{x^2(t)}} < \frac{1}{x_{cl}^2(t)} \sim |F_{cl}|.
\label{inequality_tgt0}
\end{equation}
Since the discrepancy between $\ev{x(t)}$ and $x_{cl}(t)$ keeps growing with time, $|\ev{F(x)}|$ is bound to decrease below $|F_{cl}|$ at some time. This is precisely the behavior in Fig.~\ref{fig:force}. In other words, the classical particle gets into a region close to the nucleus, where it experiences a stronger force than the quantum particle.

Here we have chosen the parameters (initial distance and momentum) in order to better illustrate some of the non-trivial features. More realistic configurations~\cite{vyas-arXiv:2011.06470} imply that there are unexpected deviations in the collision time and asymptotic trajectories.

\section{Summary and conclusions}

We have discussed the differences between the classical and quantal descriptions of the head-on collision in Rutherford's alpha-particle scattering experiment. The quantum picture assumes that the alpha particles are described by Gaussian wave packets, and the gold nucleus is represented by the Coulomb potential. We have discussed how the convexity in the repulsive Coulomb force leads to a disagreement between Hamilton's and Ehrenfest's dynamics. The deviations of $\ev{x}$ and $\ev{F}$ from their classically expected values, and the corresponding implications on the notion of `distance of closest approach' were investigated. A similar study in two-dimensions could be useful in understanding the onset of various nuclear reactions where charged projectiles need to overcome the Coulomb barrier of the target nuclei.

\section*{Acknowledgements}

This work is jointly supported by \textbf{(i)} Nanyang Technological Univesity, Singapore via NTU-India Connect Research Internship, \textbf{(ii)} Polish National Agency for Academic Exchange NAWA Project No. PPN/PPO/2018/1/00007/U/00001, and \textbf{(iii)} Indian Institute of Technology Roorkee, India via TSS-IRI grant. T.K. thanks Timothy Liew for hospitality at Nanyang Technological University, Singapore.

\bibliographystyle{apsrev4-2}
\bibliography{main}

\end{document}